\documentclass[epj]{svjour}

\usepackage{graphicx}
\usepackage{fancyhdr}
\usepackage{amsmath}
\def\makeheadbox{{%  remove EPJC header
 }}
\def\mytitle{My title} 
\def\myauthors{My name}  
\def\mytype{My type of session}
\def\mysession{My session}
\def\mytitle{Virtual Gravitons at the LHC} %Put your title here!
\def\myauthors{Tilman Plehn}    %Put your name here!
\def\mytype{Contributed Talk}    
\def\mysession{Alternatives}
\begin{document}
\title{Virtual Gravitons at the LHC}

\author{Daniel F. Litim \inst{1} \and \underline{Tilman Plehn} \inst{2}}
\institute{Department of Physics and Astronomy, University of Sussex,
  Brighton, BN1 9QH, United Kingdom \and SUPA, School of Physics, University
  of Edinburgh, United Kingdom}
\date{}
\abstract{Virtual gravitons effects at the LHC in scenarios with large
  extra dimensions and low-scale gravity are sensitive to the ultraviolet
  completion of the Kaluza-Klein effective theory.  We study implications of a
  gravitational fixed point at high energies on gravitational Drell-Yan lepton
  production in hadron collisions. The fixed point behaviour leads to finite
  LHC cross sections. We determine the reach for the fundamental Planck scale.
  An observation of these signals might shed light on the fundamental quantum
  theory of gravity.
 % end of PACS codes
} %end of abstract
\maketitle
%DO NOT REMOVE THIS LINE
%

\newcommand{\tev}{~{\ensuremath\rm TeV}}
\newcommand{\ifb}{~{\ensuremath\rm fb}^{-1}}
 
%%11.10.Hi       Renormalization group evolution of parameters
%%11.10.Jj       Asymptotic problems and properties
%%11.10.Kk       Field theories in dimensions other than four
%%04.50.+h       Gravity in more than four dimensions; 
%%11.15.Tk       Other nonperturbative techniques
%%11.25.Mj       Compactification and four-dimensional models)
%%04.60.-m       Quantum gravity
%\maketitle
 
\section{Introduction} 

Theories with large compactified extra dimensions~\cite{add} have been studied
in detail for different colliders~\cite{grw}.  In such models gravity
propagates in the higher--dimen\-sional bulk, while Standard Model particles
are typically confined to the four-dimensional brane.  If the fundamental
Planck scale $M_D$ in $(4+n)$ dimensions is indeed in the TeV range, the LHC
will is likely to see clear signals thereof.  This way the LHC becomes
sensitive to the dynamics of gravity, and could possibly become the first
experiment able to establish evidence for the quantization of gravity.
Searches for massive Kaluza--Klein (KK) gravitons at hadron colliders are
based on two signatures: real graviton emission, leading to missing transverse
momentum~\cite{grw,real_kk} and virtual graviton effects which alter the rates
and distributions of Standard Model candles like Drell--Yan or photon--pair
production~\cite{grw,virtual_kk}.  In the context of low-scale quantum
gravity, such signals have been studied in a KK effective field
theory~\cite{grw}, which allows for a controlled description as long as the
relevant momentum scales are sufficiently below an ultraviolet (UV) cutoff of
the order of the fundamental Planck scale $M_D$.\medskip

For momentum transfer near the Planck scale and above, an understanding of
gravitational interactions requires an explicit quantum theory for gravity.
It has been suggested that a local quantum theory of gravity in terms of the
metric field may very well exist on a non-perturbative level, despite its
notorious perturbative non-renormalizability~\cite{Weinberg79}.  This
``asymptotic safety'' scenario requires the existence of a non-trivial UV
fixed point for quantum gravity under the renormalization group.  In higher
dimensions, as relevant for the present setup, a new non-trivial UV fixed
point has been detected in \cite{Litim:2003vp,Fischer:2006fz}. The fixed point
implies that gravitational interactions become soft at high energies.  These
findings also support $4d$ results based on renormalization group studies in
various setups and approximations
\cite{Reuter:1996cp,Souma:1999at,Lauscher:2001rz,Reuter:2001ag,Niedermaier:2003fz,Forgacs:2002hz,Percacci:2003jz,Litim:2003vp,PT,Niedermaier:2006ns},
and lattice simulations \cite{lattice}.
With~\cite{Litim:2003vp,Fischer:2006fz} at hand, it is now feasible to
evaluate phenomenological implications for quantum gravity at the LHC
\cite{Fischer:2006fz,us,tom,frankfurt}. \medskip

Real graviton emission is weakly sensitive to the UV sector of the theory, and
an effective--field---theory approach already yields stable cross section
predictions at the LHC. This is further helped through the steep drop in the
gluon densities, which acts as an additional theory--independent UV cutoff.
Virtual graviton effects, in turn, induce Planck-scale suppressed
higher--dimensional operators which are dominated by the far UV regime of the
KK spectrum. Within effective theory, this requires an UV regularisation.
Furthermore, this strong cutoff sensitivity implies large theoretical
uncertainties on production rates at the LHC and beyond~\cite{gs,gps}.  Here,
we study the impact of a gravitational fixed--point at high energies on
virtual gravitons and lepton pair production $pp\to \ell^+\ell^-$ at the LHC
using Wilson's renormalization group \cite{us}.

\section{Gravitational fixed point} 

We first discuss implications of gravitational fixed points and
consider the renormalization group equation for the gravitational
coupling $G$ as a function of the a momentum scale $\mu$ in $D$
dimensions~\cite{Litim:2003vp,Niedermaier:2006ns}. We have
\begin{equation}\label{dg}
\beta_g\equiv
\frac{{\rm d}\, g(\mu)}{{\rm d}\ln \mu} = (D-2+\eta)g(\mu)\,,
\end{equation}
where $g(\mu)=G(\mu)\mu^{D-2}\equiv G_0 Z^{-1}(\mu)\mu^{D-2}$ is the
dimensionless gravitational coupling.  Here $\eta=-\mu\partial_\mu \ln Z$
denotes the anomalous dimension of the graviton. The wave function factor is
normalized to $Z(\mu_0)=1$ at some reference scale $\mu_0$ with $G(\mu_0)$
given by Newton's constant $G_0$. In general, the anomalous dimension depends
on all couplings of the theory. Due to its structure, eq.(\ref{dg}) predicts
two types of fixed points. At small coupling, the anomalous dimension vanishes
and $g=0$ corresponds to the non-interacting ({\it i.e.} Gaussian) fixed
point. This fixed point dominates the deep infrared region of gravity $\mu\to
0$. An interacting fixed point $g_*$ can occur if the anomalous dimension of
the graviton becomes non-perturbatively large,
\begin{equation}\label{etaUV}
\eta_* = 2-D.
\end{equation} 
Hence, a non-trivial fixed point of quantum gravity in $D>2$ implies a
negative value for the graviton anomalous dimension, precisely
counter-balancing the canonical dimension of $G$.  This means the
gravitational coupling constant scales as $G(\mu)\to g_*/\mu^{D-2}$ in
the vicinity of the non-trivial fixed point. In the UV limit
the gravitational coupling $G(\mu \to \infty)$ then becomes arbitrarily
weak.\medskip

For the explicit renormalization group equations for
gravity~\cite{Reuter:1996cp} we consider an effective action
$\Gamma_k$ with
\begin{equation}\label{EHk}
 \Gamma_k=
\frac{1}{16\pi G_k}\int d^Dx \, \sqrt{g} \, \left[-R(g)+\cdots\right]
\end{equation}
where $k$ denotes the Wilsonian renormalization--group scale replacing the
scale $\mu$ introduced in eq.(\ref{dg}), and $R(g)$ denotes the Ricci scalar.
The dots stand for the cosmological constant, higher dimensional operators in
the metric field, gravity--matter interactions, a classical gauge fixing and
ghost terms. All couplings in eq.(\ref{EHk}) become running couplings as
functions of the momentum scale $k$. In the infrared $k \ll M_D$, the
gravitational sector is well--approximated by the Einstein--Hilbert action
with $G_k\approx G_0$. The corresponding operators scale canonically. In the
UV regime $k \gg M_D$, the non-trivial renormalization--group running of
gravitational couplings becomes important. The Wilsonian
renormalization--group flow for the action eq.(\ref{EHk}) is given by an exact
differential
equation~\cite{Wetterich:1992yh,Reuter:1996cp,Litim:1998nf,Litim:2001up,Freire:2000bq}
\begin{equation}\label{ERG} 
\partial_t \Gamma_k=
\frac{1}{2} \, {\rm Tr} \left({\Gamma_k^{(2)}+R_k}\right)^{-1}\partial_t R_k
\end{equation} 
and $t=\ln k$. The trace stands for a momentum integration and a sum
over indices and fields, and $R_k(q^2)$ denotes an appropriate
infrared cutoff function at momentum scale $q^2\approx
k^2$~\cite{Litim:2001up}.\medskip

To illustrate the leading renormalization--group effects of gravity in models
with large extra dimensions and Standard--Model matter on a brane we
approximate eq.(\ref{EHk}) by the Ricci scalar and discuss the running of
$g_k$~\cite{Litim:2003vp}. The central pattern is not altered through the
inclusion of a cosmological constant \cite{Fischer:2006fz}. Using
eq.(\ref{EHk}) and eq.(\ref{ERG}), we find
\begin{equation}\label{betag0}
\beta_g=\frac{(1-4Dg)(D-2)}{1-(2D-4)g} \, g
\end{equation}
where $g$ has been rescaled by a numerical factor. Eq.(\ref{betag0})
displays a non-Gaussian fixed point at
$g_*=1/(4D)$.
Integrating eq.(\ref{betag0}), we find
\begin{equation}\label{g0-explicit}
\frac{1}{D-2}\ln\left(\frac{g_k}{g_0}\right)
-\frac{1}{\theta_{\rm NG}}\ln\left(\frac{g_*-\,g_k}{g_*-\,g_0}\right)
=
\ln\frac{k}{k_0}\,
\end{equation}
with initial condition $g_0$ at $k=k_0$, and $\theta_{\rm NG}=2D\ 
(D-2)/(D+2)$.  The result eq.(\ref{g0-explicit}) holds for generic
Wilsonian momentum cutoff, with the slight modification that the
values for $g_*$ and the scaling exponent $\theta_{\rm NG}$ can depend
on the details~\cite{Litim:2003vp,Fischer:2006fz}.  
The anomalous dimension of the graviton reads
\begin{equation}\label{eta}
\eta=\frac{2(D-2)(D+2)\,g}{2(D-2)\,g-1}\,.
\end{equation}
Inserting the 
running coupling eq.(\ref{g0-explicit}) into eq.(\ref{betag0}) 
shows that the anomalous dimension displays a smooth cross-over between the
IR domain $k\ll M_D$ where $\eta \approx 0$ and the UV domain
$k\gg M_D$ where $\eta \approx2-D$.  The cross-over regime becomes
narrower with increasing dimension~\cite{Fischer:2006fz,us}.

%********|*********|*********|*********|*********|*********|*********|****
\begin{figure*}[t]
\includegraphics[width=.95\textwidth]{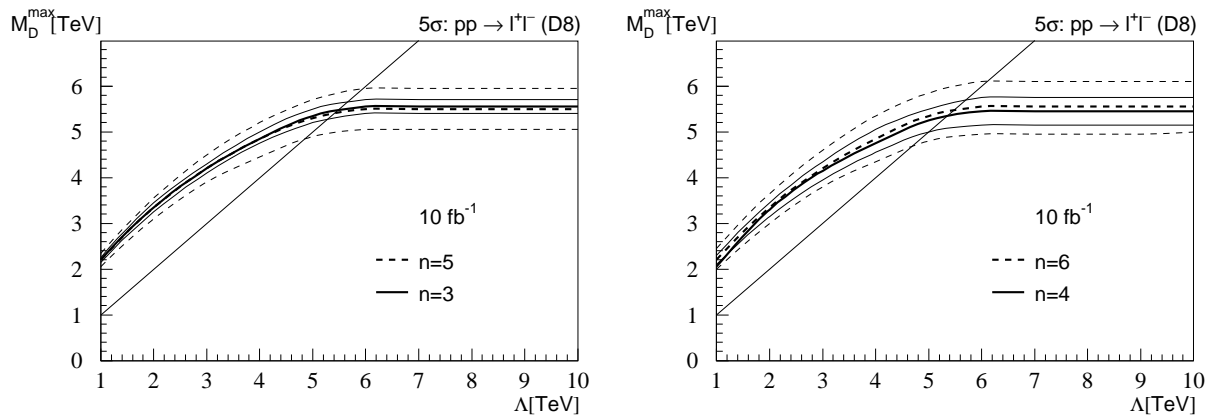}
\caption{The $5\sigma$ discovery contours in $M_D$ 
  at the LHC, shown as a function of a cutoff $\Lambda$ on $\sqrt{s} =
  E_{\rm parton}$ for an assumed integrated luminosity of $10\ifb$.
  Thin lines show a $\pm$10\% variation of $k_{\rm trans}$ about
  $M_D$, the straight line is the diagonal $M_D^{\rm max} = \Lambda$.
  The leveling-off at $M_D^{\rm max} \approx \Lambda$ reflects the
  gravitational UV fixed point.To enhance the reach we
  require $m_{\ell\ell}^{\rm min}{=}{\rm
    min}(M_D/3,2\tev)$.}
\label{fig:discovery}
\end{figure*}
%********|*********|*********|*********|*********|*********|*********|****

\section{Drell--Yan with Gravitons} 

Virtual graviton effects, as opposed to real graviton emission,
crucially depend on an UV completion, like the UV
fixed point~\cite{grw}.  Contributions to the Drell--Yan process can
be generated through a dimension--8 operator in the effective
action~\cite{grw,gs,gps}.  Tree--level graviton exchange is described
by an amplitude ${\cal A} = {\cal S}\cdot {\cal T}$, where ${\cal T} =
T_{\mu\nu} T^{\mu\nu} - T_\mu^\mu T_\nu^\nu/(2+n)$ is a function of
the energy-momentum tensor, and
\begin{equation} \label{S}
{\cal S}= \frac{S_{n-1}}{M_D^{2+n}} \; 
            \int_0^\infty d m \; m^{n-1}\, P(s,m)
\end{equation}
with $S_{n-1}=2 \pi^{n/2}/\Gamma(n/2)$ is a function of the scalar part
$P(s,m)$ of the graviton propagator~\cite{grw,gs,gps}. The integration over
the KK tower $m$ corresponds to gravity propagating in the higher-dimensional
bulk. If the graviton anomalous dimension is small, the propagator is well
approximated by the usual scalar graviton propagator as long as the relevant
momentum transfer as well as the KK masses are below the Planck scale $M_D$.
It is well known that this expression for $\mathcal{S}$ is UV
divergent for $n\ge 2$~\cite{grw}.  Implementing an UV cutoff
$\Lambda$~\cite{gs} as the upper integration boundary in the KK integration
over $m$ gives
\begin{equation}
{\cal S}_\Lambda 
         = \frac{S_{n-1}}{n-2} \frac{1}{M^4_D} \; 
           \left( \frac{\Lambda}{M_D} \right)^{n-2} \;
\left[1+{\cal O}\left(\frac{s}{\Lambda^2}\right)\right]\,.
\label{eq:s_theta}
\end{equation}
The strong cutoff dependence of $\mathcal{S}_\Lambda$ indicates that
the effective--field--theory prediction for $\mathcal{S}$ for $n\ge 2$
is indeed dominated by UV contributions and sensitive to the
UV completion of the KK theory. Note that in this approach we
apply the same cutoff to the partonic LHC energy, which means we only
evaluate contributions to $\mathcal{S}_\Lambda$ with both, $\sqrt{s} <
\Lambda$ and $m < \Lambda$.\medskip

Within asymptotically safe gravity the UV divergence in the $m$
integration is regularized by the non-trivial anomalous dimension of the
graviton. We implement this softening of gravity by evaluating the dressed
propagator $1/(Z(k^2)\ p^2)$ at momentum scale $k^2\approx p^2$ ($p^2$
denotes the relevant graviton momentum). Therefore, it scales like
$p^{-2(1-\eta(p)/2)}$, which in the far UV becomes $(p^2)^{-D/2}$.
For small $s/M^2_D$, and because of the narrow crossover of the anomalous
dimension, this amounts to the replacement
\begin{alignat}{5}
  P(s,m) = \left\{
  \begin{array}{lr}
    \displaystyle{ \frac{1}{s+m^2} }
               &m < k_{\rm trans} \\[4mm]
    \displaystyle{ \frac{k_{\rm trans}^{n+2}}{(s+m^2)^{n/2+2}} }
         \quad &m > k_{\rm trans}\,. \\
  \end{array}
                     \right.
\label{propUV}
\end{alignat}
The transition scale $k_{\rm trans}$ should be of the order of the fundamental
Planck scale $k_{\rm trans} \sim M_D$.  The integration over $m$ is finite,
and for small $s/M^2_D$ it can be performed analytically, leading to
$\mathcal{S}_{\rm FP} = \mathcal{S}_\Lambda +\mathcal{S}_{\rm UV}$.  For large
$s/M^2_D$, we implement the leading asymptotic suppression of $\mathcal{S}(s)$
by matching with $\mathcal{S}_{\rm FP}(s)$ at $s= k^2_{\rm trans}$. Because of
the steep decrease of the gluon density towards large $\sqrt{s}$ the numerical
impact of the details of this modelling can be expected to be small. For a
more detailed evaluation, see~\cite{next}.
\section{LHC Signal} 

In Fig.~\ref{fig:discovery} we display the discovery potential in
$M_D$ at the LHC. Taking into account the leading $Z$-production
background we compute the minimal signal cross section $\sigma_{\rm
  tot}(M_D)$ for which we can still observe a $5\sigma$ excess
(see~\cite{gps} for technical details).  This minimal cross section
translates into a reach $M_D^{\rm max}$.  The behavior of this
prediction as an extension of the effective--field--theory method we
check by introducing an artificial cutoff $\Lambda$ on the partonic
energy~\cite{gps}, setting $\mathcal{S}_{\rm FP}=0$ for $\sqrt{s} >
\Lambda$.  This cutoff is an unnecessary addition to our approach,
which means that for sufficiently large values, $M_D^{\rm max}$ like
any observable has to become independent of it.  This is nicely seen
in Fig.~\ref{fig:discovery}. To estimate the uncertainties in our
computation, we allow for a 10\% variation in $k_{\rm trans} \sim
M_D$, leading to mild variations in Fig.~\ref{fig:discovery} of a
similar magnitude, slightly increasing with $n$.\medskip

\begin{figure}[t]
\includegraphics[width=.45\textwidth]{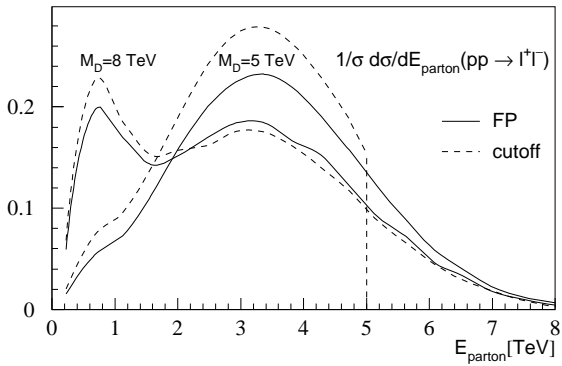}
\caption{Comparison of normalized distributions of the partonic energy 
  $E_{\rm parton}$ for the dimension--8 operator correction to Drell--Yan
  production at the LHC $(n=3)$. Full line: present work, dashed line:
  approximation eq.(\ref{eq:s_theta}) with $\Lambda=M_D$.}
\label{fig:spectrum}
\end{figure}

In Fig.~\ref{fig:spectrum} we show the normalized $\sqrt{s}$ or $E_{\rm
  parton}$ distributions for Drell--Yan production including all
Standard--Model and KK graviton contributions for $n=3$ and $M_D=5$~TeV and
$8$~TeV.  To show the entire range of $\sqrt{s}$, in contrast to
Fig.~\ref{fig:discovery} we do not apply any $m_{\ell\ell}$ cut. The solid
curves represent our fixed--point analysis. The entire $\sqrt{s}$ range
contributes to the rate as long as there is a sizeable parton luminosity and
as long as the large-$s$ suppression of $\mathcal{S}$ is not too strong. The
dashed curves correspond to the cut-off approximation $\mathcal{S}_\Lambda$
with $\Lambda =M_D$, so there is no contribution above $\sqrt{s} = \Lambda$.
The two sets of curves do not scale in a simple manner because Standard Model
and KK amplitudes interfere.  For small $\sqrt{s}$ this interference term is
significant, whereas for large $\sqrt{s}$ there is hardly any Standard--Model
contribution. Once we apply a cut of the kind $m_{\ell\ell} > M_D/3$ this
background--interference contribution will become negligible.\medskip

In Tab.~\ref{tab:rates}, we show the LHC production cross-section for
the Drell--Yan process, including virtual gravitons, for $n=3,6$. Our
fixed--point results $\mathcal{S}_{\rm FP}$ are given in $(a)$ and
$(b)$: in $(a)$, we simply retain the leading term in $s/M_D^2$ for
all values of $\sqrt{s}$.  In $(b)$, we correct the high--energy
behavior using the matched large-$\sqrt{s}$ behavior above $M_D$. In
$(c)$ we introduce a double cutoff $\Lambda = M_D$ in $m$ and
$\sqrt{s}$ into the naive KK effective theory~\cite{gps}. For large
enough $M_D\approx 5-8\tev$ we see that the LHC has little sensitivity
to quantum--gravity effects in $\sqrt{s}$, and we find only small
differences between $(a)$ and $(b)$. In that case, the difference
between $(b)$ and $(c)$ is exclusively due to the KK integration.  For
small $M_D\approx 2\tev$, all three approaches lead to significant
differences which originate from physics beyond the fundamental Planck
scale, which is omitted in the effective--field--theory approach
$(c)$.

%********|*********|*********|*********|*********|*********|*********|****
%********|*********|*********|*********|*********|*********|*********|****

\section{Conclusions} 

We have laid out a framework to study quantum--gravitational effects at high
energies within Wilson's renormalization group. This extends previous
effective--field--theory computations towards momentum regimes at and above
the fundamental Planck scale. Our approach is based on the dominant effects in
asymptotically safe gravity. It can be extended to take vertex corrections
into account~\cite{Percacci:2003jz}, in ways similar to the
systematics developed in other theories, $e.g.$ infrared QCD
\cite{Pawlowski:2003hq}.  For the physical observables studied here, we expect
vertex corrections to be subleading because the relevant momentum integrals
are dynamically suppressed above the Planck scale. \medskip 

We employed our approach to gravitational Drell-Yan production in scenarios
with large extra dimensions. The main new effects are dictated by the
gravitational UV fixed point above the fundamental Planck scale. The
renormalization--group improvement advocated here leads to finite
cross--section and to theoretically well controlled experimental signatures at
the LHC, already at low luminosities. The (model--dependent) UV contributions
to the dimension-8 operator studied here may allow to distinguish different
models for quantum gravity.

%********|*********|*********|*********|*********|*********|*********|****
\begin{table}[t]
\begin{small}
\begin{tabular}{|c|rrr|rrr|}  
   \hline
                     & & $n=3$ &           
                       & & $n=6$ &                \\[1mm]
                  & $2\tev$ & $5\tev$ & $8\tev$ 
                       & $2\tev$ & $5\tev$ & $8\tev$   \\[1mm]
  \hline
   $a$           & 2270 & 1.41 & 0.0317 &  2220 & 1.36  & 0.031 \\
   $b$ &  408 & 1.24 & 0.0317 &   398 & 1.21  & 0.031 \\
   $c$           &  173 & 0.72 & 0.0204 &    66 & 0.28  & 0.008 \\
\hline
\end{tabular}  
\end{small}
\caption{Comparison of Drell--Yan production rates at the LHC after 
  cuts for $M_D=2,5,8\tev$.  See main text for the definitions of 
  the scenarios $(a)$, $(b)$ and $(c)$.} 
\label{tab:rates}
\end{table}
%********|*********|*********|*********|*********|*********|*********|****

\end{document}